\documentclass[12pt]{article}

\usepackage[letterpaper,hmargin=1in,vmargin=1in]{geometry}

\usepackage{graphicx,epstopdf,amsmath,amsfonts}

\def\be{\begin{equation}}
\def\ee{\end{equation}}
\def\ba{\begin{eqnarray}}
\def\ea{\end{eqnarray}}
\def\ge{\mathrel{\raise.3ex\hbox{$>$\kern-.75em\lower1ex\hbox{$\sim$}}}}
\def\la{\mathrel{\raise.3ex\hbox{$<$\kern-.75em\lower1ex\hbox{$\sim$}}}}

\def\simgt{\mathrel{\raise.3ex\hbox{$>$\kern-.75em\lower1ex\hbox{$\sim$}}}}
\def\simlt{\mathrel{\raise.3ex\hbox{$<$\kern-.75em\lower1ex\hbox{$\sim$}}}}

\newcommand{\bi}[1]{\bibitem{#1}}
\newcommand{\fr}[2]{\frac{#1}{#2}}

\newcommand{\nc}{\newcommand}

\nc{\gone}{\bar g_{\pi NN}^{(1)}}
\nc{\gzero}{\bar g_{\pi NN}^{(0)}}
\nc{\al}{\alpha}
\nc{\ga}{\gamma}
\nc{\de}{\delta}
\nc{\ep}{\epsilon}
\nc{\ze}{\zeta}
\nc{\et}{\eta}
\nc{\ka}{\kappa}
\nc{\rh}{\rho}
\nc{\si}{\sigma}
\nc{\ta}{\tau}
\nc{\up}{\upsilon}
\nc{\ph}{\phi}
\nc{\ch}{\chi}
\nc{\ps}{\psi}
\nc{\om}{\omega}
\nc{\Ga}{\Gamma}
\nc{\De}{\Delta}
\nc{\La}{\Lambda}
\nc{\Si}{\Sigma}
\nc{\Up}{\Upsilon}
\nc{\Ph}{\Phi}
\nc{\Ps}{\Psi}
\nc{\Om}{\Omega}
\nc{\ptl}{\partial}
\nc{\del}{\nabla}
\nc{\ov}{\overline}
\nc{\newcaption}[1]{\centerline{\parbox{15cm}{\caption{#1}}}}

\def\beq{\begin{equation}}
\def\eeq{\end{equation}}
\def\bmat{\begin{displaymath}}
\def\emat{\end{displaymath}}
\def\bear{\begin{eqnarray}}
\def\eear{\end{eqnarray}}
\def\ba{\begin{eqnarray}}
\def\ea{\end{eqnarray}}
\def\bery{\begin{array}}
\def\ery{\end{array}}
\def\bit{\begin{itemize}}
\def\eit{\end{itemize}}
\def\ben{\begin{enumerate}}
\def\een{\end{enumerate}}
\def\btab{\begin{tabular}}
\def\etab{\end{tabular}}
\def\btbl{\begin{table}}
\def\etbl{\end{table}}
\def\bfig{\begin{figure}[htb]}
\def\efig{\end{figure}}
\def\bpic{\begin{picture}}
\def\epic{\end{picture}}

\def\ga{\mathrel{\raise.3ex\hbox{$>$\kern-.75em\lower1ex\hbox{$\sim$}}}}
\def\la{\mathrel{\raise.3ex\hbox{$<$\kern-.75em\lower1ex\hbox{$\sim$}}}}
\def\gappeq{\mathrel{\rlap {\raise.5ex\hbox{$>$}}
{\lower.5ex\hbox{$\sim$}}}}
\def\lappeq{\mathrel{\rlap{\raise.5ex\hbox{$<$}}
{\lower.5ex\hbox{$\sim$}}}}

\def\gyr{{\rm \, G\kern-0.125em yr}}
\def\mev{{\rm \, Me\kern-0.125em V}}
\def\gev{{\rm \, Ge\kern-0.125em V}}
\def\tev{{\rm \, Te\kern-0.125em V}}

\begin{document}

\begin{titlepage}

\rightline{FTPI--MINN--08/29}
\rightline{UMN--TH--2709/08}

\setcounter{page}{1}

\vspace*{0.2in}

\begin{center}

\hspace*{-0.6cm}\parbox{17.5cm}{\Large \bf \begin{center}

Bosonic super-WIMPs as keV-scale dark matter
\end{center}}

\vspace*{0.5cm}
\normalsize

\vspace*{0.5cm}
\normalsize

{\bf  Maxim Pospelov$^{\,(a,b)}$, Adam Ritz$^{\,(a)}$ and Mikhail Voloshin$^{\,(c,d)}$}

\smallskip
\medskip

$^{\,(a)}${\it Department of Physics and Astronomy, University of Victoria, \\
     Victoria, BC, V8P 1A1 Canada}

$^{\,(b)}${\it Perimeter Institute for Theoretical Physics, Waterloo,
ON, N2J 2W9, Canada}

$^{\,(c)}${\it William I.\ Fine Theoretical Physics Institute,\\
University of Minnesota, Minneapolis, MN~55455, USA}

$^{\,(d)}${\it Institute of Theoretical and Experimental Physics, Moscow, 117218, Russia}

\smallskip
\end{center}
\vskip0.2in

\centerline{\large\bf Abstract}

We consider models of light super-weakly interacting cold dark matter, with ${\cal O}(10-100)$ keV mass,
focusing on bosonic candidates such as pseudoscalars and vectors. 
We analyze the cosmological abundance, the $\gamma$-background created by particle
decays, the impact on stellar processes due to cooling, and the direct detection 
capabilities in order to identify classes of models that pass
all the constraints. In certain models, variants of photoelectric (or axioelectric) absorption 
of dark matter in direct-detection experiments can provide a sensitivity to the superweak couplings 
to the Standard Model which is superior to all existing indirect constraints. In all models studied, the annual modulation of
the direct-detection signal is at the currently unobservable level of  $O(10^{-5}$).

\vfil
\leftline{July 2008}

\end{titlepage}

\section{Introduction}

The evidence for the existence of non-baryonic dark matter (DM) now comes from many sources and ranges over 
many distance scales \cite{review}, from the rotation curves of galaxies, the dynamics of clusters, lensing data and the characteristics of
large-scale structure, to the features of the cosmic microwave background (CMB) fluctuation spectrum and the success of 
big bang nucleosynthesis (BBN). All of these pieces of astronomical data point to a similar cosmological density of 
dark matter, several times that of visible baryonic matter. However, this data only probes the gravitational interaction 
of dark matter and, while it presents us with one of the most compelling arguments for physics beyond the Standard Model,
gaining insight into its non-gravitational interactions remains a primary experimental focus, both through underground detectors, 
particle colliders, and the observation of photon and neutrino fluxes from overdense regions in the galaxy and beyond.

This lack of any direct information on how dark matter may couple to the Standard Model (SM) means we are forced to rely on various
theoretical expectations. In particular, the many successes of standard cosmology motivate a simple 
thermal mechanism for populating the universe with dark matter with a well-defined freeze-out abundance as the universe expands. This 
in turn requires a specific annihilation cross-section which is necessarily non-gravitational in origin. 
The fact that a weakly interacting particle with a weak-scale mass has an annihilation rate in the right ballpark \cite{lw,review},
combined with our expectations for new physics at the electroweak scale, has rightly led to
the prevailing WIMP paradigm for cold dark matter.
Nonetheless, persistent problems in understanding the small-scale gravitational clustering 
properties in cold dark matter simulations and on galactic scales has motivated variants
of this picture where the dark matter may be somewhat lighter, with masses down to the keV range. 
Masses in this range imply a super-weak interaction strength between 
dark matter and the SM sector, indeed many orders of mangitude below 
weak-scale cross sections. This follows from the necessity to have
early thermal decoupling of the DM sector, prior to the electroweak epoch at $T\sim 100$ GeV,
in order to satisfy the conflicting requirements of not having too much energy density in 
dark matter, and the strong lower bounds on $m_{DM}$ coming from the analysis of 
structure formation. If early decoupling can be achieved, 
then masses in the keV range may withstand these combined constraints and at the same time provide a 
rather attractive mechanism for ensuring the correct dark matter energy density.

An important feature of keV-scale dark matter is that, unlike the majority of electroweak scale WIMP models, it need not be stable
against decays to light SM degrees of freedom,
{\em e.g.} photons and neutrinos. 
Given the super-weak strength of its interaction such decays
may be strongly inhibited, but nevertheless keV dark matter can be emitted or absorbed in 
astrophysical environments and in terrestrial experiments, and thus is 
possibly subject to additional constraints.  
The main question to be studied in this paper  is  whether the application of 
cosmological and astrophysical constraints leaves any  realistic chance for
the direct detection of such keV-scale candidates. The best studied models in this 
class are the sterile neutrino and gravitino, examples of  fermionic superWIMPs \cite{sterile,swimp}, 
where it is well-known that there are no chances for 
direct detection. In contrast, the present paper studies
bosonic superWIMP models, which need not entail such a pessimistic conclusion. In particular, we show that the 
potential sensitivity of direct detection experiments 
to bosonic keV dark matter can indeed be competitive with astrophysical bounds as well as 
 with the lifetime and cosmic gamma-background constraints. 

The majority of underground direct-detection experiments are specifically tuned to the 
nuclear recoil energy regime 
 and all $\gamma$- or $\beta$-like events are typically 
considered as background. 
 Currently, the experiments with the best limits are now
sensitive to picobarn-scale cross-sections for WIMP-nucleus elastic scattering, when the WIMP mass
is on the order of the electroweak scale. 
To illustrate our main point, we  present a simple estimate that compares the rate of absorption of  10~keV-mass axion-like
dark matter ($a$-particles),  coupled to electrons with strength $m_e/f_a 
=m_e(10^{10}~{\rm GeV})^{-1}$, with the recoil
signal from a $1$ TeV WIMP that has an elastic scattering cross section on the  nucleus  of $\sigma_{el}
\sim 10^{-36}$cm$^2$:
\be
\fr{\sigma_{abs} v_{\rm DM}n_a}{\sigma_{el} v_{\rm DM}n_{\rm WIMP}}\sim \fr{m_{\rm WIMP} }{m_a } 
\times  \fr{c}{v_{\rm DM}}
\times \fr{(10^{10}~{\rm GeV})^{-2}}{10^{36}~{\rm cm}^2} \sim 10^{8}\times 10^{3}\times 10^{-11} \sim O(1).
\label{estimate}
\ee
For this rather crude estimate we simply took $\sigma_{abs} v_{\rm DM} \sim f_a^{-2}$, and 
$m_an_a\sim m_{\rm WIMP}n_{\rm WIMP}\sim \rho_{DM}$. 
Although the superWIMP absorption cross section is  orders of magnitude below the 
weak cross section, it is compensated by the tremendous gain in the local number 
density of dark matter particles, and by the fact that the inelastic cross section 
$\sigma_{abs}$ scales inversely with the dark matter velocity. Although rather imprecise, the estimate (\ref{estimate}) illustrates that the absorption of 
10 keV dark matter can indeed produce signals that are well within modern detection 
capabilities. Moreover, the scale $f_a \sim 10^{10}$ GeV is at or above the limits imposed by even the most 
stringent constraints on 
star cooling. We believe that this is an important point, and the 
sensitivity to new and viable dark matter scenarios  can (and should) be explored by direct detection experiments. 
For comparison, fermionic candidates in the superWIMP class, a well-studied example 
of which is sterile neutrino dark matter, do not fall into the class covered by the estimate (\ref{estimate})
for many different reasons; the primary one being that sterile neutrinos are not 
fully absorbed but rather converted to active neutrinos that carry away most of the 
rest energy.

Thus far, the issue of direct detection of keV dark matter has only been discussed by the
DAMA collaboration \cite{DAMA1}\footnote{As this paper was being 
prepared for publication, another 
experimental collaboration, CoGeNT \cite{cogent}, reported their results on the 
absorption of keV-scale superWIMPs.} in connection with an annual modulation signal initially 
reported by DAMA/NaI \cite{DAMA3} and recently confirmed by DAMA/Libra \cite{DAMA2}. If indeed the annual modulation 
signal of DAMA were to be attributed to the absorption of keV dark matter,\footnote{See \cite{dama_wimp,cogent} for recent analyses of
WIMP models in relation to the DAMA results.} which would 
cause ionization but no significant recoil, it could provide a plausible explanation for 
why other front-running experiments such as CDMS and Xenon \cite{cdms,xe} see no signal. 
This was the main point of Ref.~\cite{DAMA1}, and in this paper we provide a re-analysis 
of this possibility, reaching instead a negative conclusion: the models that were presented in \cite{DAMA1}
as explaining the modulation are in fact ruled out either by lifetime arguments, by 
astrophysical constraints, or directly by the large unmodulated counting rates in 
underground detectors.

To make the discussion sufficiently general, we focus on three generic possibilities for bosonic superWIMP dark matter:
pseudoscalars, scalars and vectors, to be defined more precisely in Section~2. Some of these models 
are technically natural in the sense of having protection for the light  dark matter mass, 
either by symmetry alone or by symmetries combined with the imposed smallness of their coupling to the 
visible sector. For each of these models, in Section~3 we analyze the lifetime, emission, and absorption rates 
that determine the dark matter abundance, the 
level of diffuse and galactic gamma backgrounds, 
the efficiency of star cooling, and the rates for direct detection. The appropriate parameter space for each model
is considered in Section~4, where we impose the relevant astrophysical constraints and determine
the viability of direct detection. We also comment on the low level of any modulated component of
the signal, and conclude with some additional remarks in Section~5.

\section{Light dark matter candidates}

In this section we list the dark matter candidate scenarios to be considered. 

\bigskip
\noindent$\bullet$ {\it Pseudoscalar DM}

We start with the
pseudoscalars $a$, and write the interactions as a combination of several derivative-like
operators of dimension five: 
\be
 {\cal L}_{\rm int} = \frac{C_\gamma a}{f_a} F_{\mu\nu}\tilde{F}^{\mu\nu} -  
\fr{\ptl_\mu a}{f_a} \bar\ps \gamma^\mu\gamma^5\ps + \cdots 
 \label{ps}
\ee
where $F_{\mu\nu}$ and $\psi$ are the electromagnetic field strength and the Dirac field of the electrons, 
and the ellipsis denotes possible interactions with other fermions and  gauge bosons, and for simplicity 
we shall assume a similar strength for the $a$-SM couplings in those sectors. 
Notice that the other possible pseudoscalar coupling to the electron, $ a \bar\ps i\gamma^5 \ps $,
can always be decomposed into the two operators in (\ref{ps}) using the equations of motion once
we account for the chiral anomaly. 
While the dimensionful coupling $f_a$ does regulate the overall strength of the SM-$a$ interaction, 
the dimensionless coupling to photons $C_\gamma$ is crucial for determining the 
lifetime and $\gamma$-background created by $a$ decay. Restricting ourselves to the
electron-photon sector, we expect three generic possibilities 
for the size of the coupling $C_\gamma$:
\begin{eqnarray}
&{\rm A:}&~~~C_\gamma \sim \fr{\pi}{\alpha}  \nonumber
\\
&{\rm B:}&~~~C_\gamma =\fr{\alpha}{4\pi}  
\\
&{\rm C:}&~~~ C_\gamma \sim \fr{\alpha}{\pi}\times \fr{m_a^2}{m_e^2} \nonumber
\nonumber
\end{eqnarray}
Case A corresponds to a pseudoscalar coupled to photons at some UV normalization scale, 
with couplings to electrons generated radiatively. Having normalized the electron coupling to $1/f_a$ in (\ref{ps}), the 
coupling $C_\gamma \gg 1$ in this case. Case C is the inverse of Case A. The derivative 
coupling to the electron axial-vector current may only lead to the $F_{\mu\nu}\tilde{F}^{\mu\nu}\ptl^2a$
operator at loop level, hence the $(m_a/m_e)^2$ suppression. Finally, Case B is intermediate, when 
$a$ is initially coupled to the fermion via the $m\psi i\gamma_5\psi$  pseudoscalar operator.
Clearly any of these three choices can be realized without fine tuning 
given an appropriate UV completion. An additional advantage of this model is the 
automatic protection of the pseudoscalar mass against radiative corrections, 
exactly as in the conventional axion case. As we are going to see later, only option 
C allows for the possibility of keV-scale dark matter without imposing 
overly strong constraints on the size of $f_a$.

\bigskip
\noindent$\bullet$ {\it Scalar DM}

A similar looking Lagrangian can be written for the scalar case:
\be
 {\cal L}_{\rm int} = \frac{C_\gamma s}{f_s} F_{\mu\nu}{F}^{\mu\nu} -  
\fr{ s}{f_s} m_e\bar\ps \ps + \cdots
 \label{s}
\ee
Here there is clearly no protection for the mass against radiative corrections. 
However, one can still exploit the smallness of the coupling $f_s^{-1}$, to render a keV-scale
mass technically natural. A one-loop correction will typically induce a mass term
that scales as 
\be
\label{naturalness}
\Delta (m_s)^2 \sim \fr{m_f^2\Lambda_{\rm UV}^2}{f_s^2},
\ee
where $m_f$ is the mass of the heaviest fermion, and $\Lambda_{\rm UV}$ is the ultraviolet cutoff. 
Taking both to the weak scale (implying supersymmetry), and requiring $m_s\sim 10 $ keV 
is equivalent to having $f_s \ga 10^9$ GeV, in other words right at the boundary of the interesting regime 
for the couplings. As for the couplings to photons, both Cases A and B cases are plausible, while 
case C is tricky and requires some fine-tuned UV physics to cancel the main contribution from the 
$m_e$ threshold. Finally, we note that the simplest renormalizable and SM-gauge-invariant realization of 
(\ref{s}) is to have the scalar singlet $s$ coupled to the Higgs doublet via the 
relevant operator $sH^\dagger H$. We will not consider the scalar example in detail in what follows, but in many ways
its phenomenology is similar to that of the pseudoscalar case.

\bigskip
\noindent$\bullet$ {\it Vector DM}

Finally, we introduce a model of keV-scale  vector dark matter. We choose the intitial 
Lagrangian in the form identical to that studied in \cite{PRV}, where an extra U(1)$'$ 
gauge field is coupled to the SM via kinetic 
mixing with the hypercharge field strength,
\be
{\cal L} = -\fr{1}{4}V_{\mu\nu}^2 - 
\fr{\kappa}{2}  V_{\mu\nu} F_{\mu\nu} + {\cal L}_{h'} + {\cal L}_{\rm dim > 4},
\label{VF}
\ee
with ${\cal L}_{h'}$ encoding the physics responsible for breaking the U(1)$'$, and 
${\cal L}_{\rm dim > 4} $ includes possible non-renormalizable higher-dimension interaction terms 
such as $H^\dagger H  V_{\mu\nu}^2$, $V_{\mu\nu}^2 F_{\alpha\beta}^2$, etc.
 After the breaking of this secluded U(1)$'$, 
the model takes the simplest possible form,
\be
{\cal L} = -\fr{1}{4}V_{\mu\nu}^2 +\fr{1}{2} m_V^2 V_\mu^2 + \kappa V_\nu \partial_\mu F_{\mu\nu},
 +\cdots,
\label{VdB}
\ee
where we retained only relevant and marginal operators, and suppressed the U(1)$'$ Higgs sector. 
This is one of the simplest UV-complete extensions of the SM, and it has been addressed in
connection with electroweak-scale  physics on a number of occasions \cite{Uprime,PRV}. Most recently 
a study of cosmology in this model was performed for small sub-eV values of $m_V$ \cite{Ringwald}. 
For vanishingly small values of $m_V^2$, the extra sector decouples as mixing with the 
photon can be reabsorbed into the mass term. This leads to an additional suppression of 
$\gamma$-$V$ conversion at temperatures much in excess of $m_V$ \cite{Ringwald}, but for
keV-scale dark matter this issue is often less important, and (\ref{VdB}) can simply be traded
for 
\be
{\cal L} = -\fr{1}{4}V_{\mu\nu}^2 +\fr{1}{2} m_V^2 V_\mu^2 + e\kappa V_\nu \ps \gamma_\mu \ps
 +\cdots,
\label{Vj}
\ee
where $V_\mu$ couples to the electromagnetic current. 
For convenience, we introduce an analogue of the electromagnetic coupling and of the fine structure constant, 
\be
e'=e \kappa,~~~~~\alpha' = \fr{(e\kappa)^2}{4 \pi},
\ee
that necessarily appear in all rates for the emission or absorption of U(1)$'$ vectors 
by SM particles.

Obviously, the mass of the $U'(1)$ gauge boson is protected by gauge symmetry, and the 
question of naturalness is relegated to the corresponding U(1)$'$ Higgs sector. However, given the smallness
of the couplings that we are going to consider, {\em e.g.} $\kappa\sim O(10^{-10})$, the 
naturalness problem is no more severe than for the SM Higgs. 

The coupling of $V$ to neutrinos is also possible via mixing with 
the $Z$-boson. This mixing, however, is further suppressed by a factor of
$(m_a/M_Z)^2 \sim 10^{-14}$, and we will disregard it in the
analysis of the model. Other realizations of U(1)$'$ \cite{Fayet,Langacker}, such as a 
gauged version of $B-L$, would allow couplings to neutrinos 
of the same size as the couplings to electrons.

\section{Decay, emission and absorption of superWIMPs}

\subsection{Pseudoscalar (and scalar) DM}

{\it Decays}: 
Given an $a$-boson (or $s$-boson) mass below the electron threshold, the decays 
will be almost exclusively to photons, and mediated by the 2-photon
interaction in (\ref{ps}), with the appropriate 
low energy value of $f_a$. The decay is shown in Fig.~1 and the width is
\be
 \Ga_{a\rightarrow 2\gamma} = \frac{C_\gamma^2}{4\pi f_a^2} m_a^3.
\ee

\begin{figure}
\centerline{\includegraphics[bb=0 580 600 730, clip=true, width=10cm]{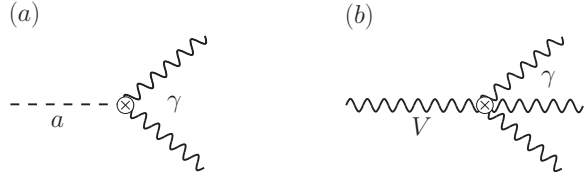}}
 \caption{\footnotesize Dominant decays to photons. (a) 
2-photon decay of the pseudoscalar $a$, and (b) the 3-photon decay of the 
 vector $V$. }
\label{f1} 
\end{figure}

Requiring that the dark matter lifetime be at least the age of the universe implies,
\be
\tau_{\rm U} \Ga_{a\rightarrow 2\gamma} \la 1 ~~\Longrightarrow ~~ 
C_\gamma^2 \leq 2 \times 10^{-6} \times \left(\frac{f_a}{10^{10}\, {\rm GeV}}\right)^2 \times \left(\frac{10\, {\rm keV}}{m_a}\right)^3,
\ee
which we see is already a significant constraint for Case A in particular, and also for 
Case B if we stick with fiducial values for $f_a$ and $m_a$.
In constrast, Case C is less constrained. However, we also need to 
consider the $\gamma$-background created by $a$-decays and, as we will
see in the next section, this provides a more stringent constraint (as is also the
case for majoron models of dark matter, discussed recently  in  \cite{valle}).

\noindent{\it Emission}: 
An important process to consider is the $a$-emission from  thermal states. 
This is relevant for determining the relic cosmological 
abundance, but is also the source of important constraints arising 
from new energy-loss mechanisms in stars. To obtain an estimate
of the impact on star-cooling, we will focus on the 
Compton-like process $e+\gamma \rightarrow e + a$. 
Working in the limit $m_a, \omega \ll m_e$, we obtain the following cross section, 
\be
\label{full}
\si_{e\gamma \rightarrow ea} = \alpha\frac{\om^2v_a}{m_e^2f_a^2}\times 
\left[ \left(1+\fr{v_a^2}{3}\right)
\left(1+\fr{m_a^2}{2\omega^2}\right) -\fr{m_a^2}{\omega^2}\left(1-\fr{m_a^2}{2\omega^2}\right)    \right]  ,
\ee
where $\omega$ is the photon frequency, and $v_a=(1-m_a^2/\omega^2)^{1/2}$ is the velocity of the 
outgoing massive axion. 
In the limit $m_a\to 0 $, $v_a \to 1$, it reduces to a well-known result in conventional axion physics 
(See, {\em e.g.} \cite{rw} and references therein):
\be
\label{reduced}
 \si_{e\gamma \rightarrow ea} = \frac{4\al}{3} \frac{\om^2}{m_e^2f_a^2}.
\ee 

The cross sections (\ref{full}) and (\ref{reduced}) for pseudoscalar production can be  translated to an energy-loss flux (energy/volume/time)
relevant for solar and red giant physics. In the limit of small axion mass  we can estimate 
this rate as 
\be
 \Ph_{e\gamma\rightarrow ea} = n_\gamma n_e \langle \om \si_{e\gamma \rightarrow ea}\rangle = \left(\frac{2\zeta(3)}{\pi^2}T^3\right) 
 \left(\frac{p_F^3}{3\pi^2}\right)\left(\frac{16\pi^6\al}{189\zeta(3)} \frac{T^3}{m_e^2f_a^2}\right),
\ee
where $p_F$ is the electron Fermi momentum. For $m_a> T$ this formula needs to be supplemented by a factor of 
$\exp(-m_a/T)$ to account for the Boltzmann-suppressed fraction of photons with energies 
above the axio-production threshold.  We will compare this flux to various constraints on stellar energetics in the next section.

One of the most stringent astrophysical constraints on exotic particles often comes from
supernova (SN) physics. Owing to the high temperature scale $\sim O(10~{\rm MeV})$ 
during the explosion,  the coupling to electrons is not the dominant 
mechanism for axion production, as the rate effectively receives an additional suppression by a factor of 
$m_e^2/T^2$. It is well known that the coupling of $a$ to nucleons provides far better sensitivity \cite{raffelt},
and for the purpose of making an estimate we shall explore $f_{aqq} \sim f_{aee} \sim f_a$. Since we consider  
pseudoscalars in the keV mass range, $T_{\rm SN} \gg m_a$, and the emission of axionic dark matter particles in supernovae then differs little 
from the standard case of ``invisible" axions \cite{raffelt}. 

The same argument applies to the thermal emission of axions in the early Universe. 
This occurs due to the interaction of SM fermions in the primordial plasma, and 
scales as $\Gamma_{a\psi} \sim Tm_\psi^2f_{a\psi\psi}^{-2}$ as long as the fermionic species $\psi$ is present,
$T\ga m_\psi$,  and $\ps$ couples to the axion with a dimensionless coefficient $\sim m_\ps/f_{a\psi\psi}$. 
Taking into account that the Hubble rate scales as $H \sim N^{1/2}T^2 M_{\rm Pl}^{-1}$,
where $N$ is the effective number of degrees of freedom, we can estimate 
the resulting number density of pseudoscalar particles weighted by the entropy,
\be
\label{cosmostuff}
\fr{n_{a}}{s} \sim \sum_\psi \int \fr{\Gamma_{a\psi}dT}{NHT} \sim 
\sum_\psi B_\psi \fr{M_{\rm Pl}m_\psi}{N^{3/2}f^2_{a\psi\psi}},
\ee
where in the last relation we took into account that $n_{a}/{s}$ is maximized near the 
annihilation/decay threshold, $T\sim m_\psi$, and the numerical constants 
$B_{\psi}$ are introduced to account for specific details of each $\psi$ threshold. 
It was also assumed that pseudoscalar production happens with a rate slower than the 
Hubble expansion, so that  the $a$-bosons are never fully thermal. 
Assuming no hierarchies among the couplings to the SM 
fermions, one observes that the largest contributions come from the heaviest fermions,
which presumably will be the top and bottom quark. In principle, all $B_\psi$ parameters can be calculated exactly, 
but we will not pursue this here as it will take us too far afield and the resulting formulae will still 
contain the model-dependent factors $f_{aqq}$. Instead, we will restrict ourselves to a simple dimensional 
estimate of the dark matter axion abundance produced via the $b$-quark (production from the top 
quark depends on other details such as the electroweak phase transition):
\begin{eqnarray}
\label{Omegaa}
\fr{\Omega_{a}}{\Omega_{\rm baryon}} \simeq \fr{m_a}{1~ {\rm GeV}} \times \fr{n_a}{s} \times \fr{s}{n_b}
\sim 5 B_b \times \fr{m_a}{\rm keV} \times \left( \fr{10^{10}~{\rm GeV}}{f_{abb}}\right)^2.
\end{eqnarray}
Since $B_b$ is naturally of $O(0.1-1)$, one can see that thermally generated peudoscalar 
dark matter has an abundance in the right ballpark given a keV-scale mass and 
the most interesting range for the coupling, 
$f_{abb} \sim 10^{10}$ GeV. A precisely analogous argument holds for 
the scalar super-WIMP abundance, where a coupling scale in excess of $10^{10}$ GeV 
is further supported by the technical 
naturalness argument, Eq. (\ref{naturalness}).

\noindent{\it Absorption}:
For the energy range considered here, the axioelectric variant of the photoelectric effect is the most
important process. 
A calculation of the axioelectric effect on an atom was performed earlier
\cite{Dimopoulos} in connection with the possibility of detecting axions with keV energy
emitted by the solar interior. We would like to rederive the 
same effect in our setting, specializing to absorption of the nonrelativistic keV mass pseudoscalar. 
Working to leading order in $v_{a}/c$, 
one can reduce the interaction in (\ref{ps}) to the following term in the 
nonrelativistic electron Hamiltonian:
\be
\label{Hint}
{\cal L}_{\rm int} = -\fr{\ptl_\mu a}{f_a} \bar\ps \gamma^\mu\gamma^5\ps ~~~
\Longrightarrow ~~~ H_{\rm int} =  \fr{\ptl_t a}{f_a} \fr{( {\bf p} \mbox{\boldmath $\sigma$})}{m_e},
\ee
where ${\bf p}$ is the momentum operator for the electron. 
Using $a(t)\sim \exp(-im_at)$ and the nonrelativistic Hamiltonian, the 
matrix element of $H_{\rm int}$ reduces to the following expression, 
\be
M_{fi} = \fr{m_a^2}{f_a} \langle f | ({\bf r} \mbox{\boldmath $\sigma$}) |i\rangle,
\label{mat_el}
\ee
where $i$ and $f$ are initial and final state electron wave functions
with $E_f-E_i = m_a$. 
This is  analogous to the amplitude for the $E1$ absorption of a photon, 
$M_{fi}^{(\gamma)} = \omega \langle f | e({\bf r} \mbox{\boldmath $\epsilon$}) |i\rangle$, with the 
photon polarization $\mbox{\boldmath $\epsilon$}$ exchanged for the spin operator. 
Thus the axioelectric effect is very similar to the photoelectric effect 
with photon energy $\omega = m_a$.
The only difference is that the wave function of the 
absorbed photon contains a space-dependent factor $\exp(i{\bf kr})$ with 
$k = \omega$, which in the 
absorption of a massive vector particle is replaced by $\exp(im_a{\bf vr})$.
Here ${\bf v}$ is the velocity of the incoming DM particle. Since $v\sim 10^{-3}c$, the latter 
oscillating  factor can safely be taken to 1. The effect of the photon spatial momentum is 
however parametrically small in comparison with the momentum transfer to the final electron, 
$O(\sqrt{m_e \, m_a})$, as long as the energy in these processes is much smaller than the electron mass $m_e$.
Neglecting the minor effect of the photon spatial momentum, squaring (\ref{mat_el}),
averaging over the initial spin, and summing over the final, 
we arrive at the following approximate relation between the cross sections for 
axion absorption and the photoelectric effect:
\be
\label{sigmaa}
\frac{\sigma_{abs} v}{\sigma_{photo}({\omega=m_a})c} ~\simeq~ \frac{3 \, m_a^2}{4 \pi \, \alpha \, f_a^2 }.
\ee
Notice that ratio (\ref{sigmaa}) remains finite even in the limit of $v \to 0$. 

For the purposes of an experimental search for pseudoscalar dark matter, it is 
useful to express (\ref{sigmaa}) directly as a counting rate in a detector consisting of a 
single atomic species of atomic mass $A$,
\be
\label{countinga}
R \simeq   \frac{1.2 \times  10^{19}}{A} g_{aee}^2 \left( \fr{m_a}{\rm keV} \right) \left(\fr{\sigma_{photo}}{\rm bn}\right) \;{\rm kg^{-1}day^{-1}},
\ee
where we used the local dark matter density, $\rho_{DM} = 0.3 ~{\rm GeV~ cm^{-3}}$, and 
introduced the dimensionless coupling $g_{aee} = 2m_e/f_a$ to allow for direct comparison with the existing results
of DAMA \cite{DAMA1,DAMA2} and the most recent experimental paper by the CoGeNT colaboration \cite{cogent}.

Note that our results differ from those quoted in \cite{DAMA1} and \cite{cogent}.\footnote{Following the release of version~1 of the present paper, the 
rate formula used in \cite{cogent} -- which originally made use of the relativistic expression for $\si_{abs}$ of \cite{Dimopoulos} --
was corrected in 0807.0879v4 to use the appropriate non-relativistic expression (\ref{sigmaa}), with a result for $R$ consistent with (\ref{countinga}).} 
 We observe that the expression used in \cite{DAMA1} for the 
axio-ionization cross section results from the omission of the leading term 
in the axion-electron Hamiltonian (\ref{Hint}). Indeed, Eq.~(45) of Ref.~\cite{DAMA1}
has an interaction term that vanishes as $v_{DM}\to 0$ giving a 
subleading contribution to the counting rate suppressed by $v_{DM}^2\sim 10^{-6}$.
We shall return to this issue in section 4.2. 

For completeness,  we will also quote 
the  axioelectric  cross section for massless axions with energy of ${\cal O}$(keV), as this result is
in widespead use for the solar physics of axions.  Taking $\omega_a \ll m_e$, and $m_a \to 0$, we arrive at the 
following analogue of the matrix element (\ref{mat_el}),
\be
M_{fi} = \fr{\omega_a^2}{f_a} \langle f | ({\bf r} \mbox{\boldmath $\sigma$}) - 
({\bf n} \mbox{\boldmath $\sigma$})({\bf n} {\bf r})|i\rangle,
\label{mat_el_rel}
\ee
where ${\bf n}$ is the direction of the incoming axion. Following the same route as before, 
we arrive at a relation between the photo- and axio-electric cross sections: 
\be
\left. \frac{\sigma_{abs} }{\sigma_{photo}({\omega=\omega_a})}\right|_{m_a\rightarrow 0} ~
\simeq~ \frac{ \omega_a^2}{2 \pi \, \alpha \, f_a^2 }.
\ee
We observe that this result is twice as large as the formula regularly quoted in the
literature (see {\em e.g.} \cite{Dimopoulos}). The source of the discrepancy with previous calculations 
can be traced to the absence of the first term in Eq.~(\ref{mat_el_rel}), and this
corrected formula may prove useful for solar axion searches.

Finally, on a more pedagogical note, we would like to demonstrate explicitly that the alternative
choice for the pseudoscalar coupling to electrons, $(2m_e/{f_a}) a \bar\ps i\gamma^5\ps$ in 
${\cal L}_{\rm int}$, leads to the same expression for $H_{\rm int}$ and the matrix elements, as the axial vector 
coupling (\ref{Hint}). Since both forms are related up to the total derivative, this must necessarily be the
case, but it is useful to see how this works in detail. We will only consider the case of a massive
nonrelativistic pseudoscalar as is relevant for this paper and, using the $v/c$ expansion, we can then write
the lower component $\chi({\bf r})$ of the Dirac spinor $\psi$ in terms of the 
upper component $\phi({\bf r})$, the total energy $E$, the potential energy $U({\bf r})$ and the helicity operator 
$( {\bf p} \mbox{\boldmath $\sigma$})$,
\be
2m_e\chi({\bf r}) = \left(1 - \fr{E-U({\bf r})}{2m_e}\right)( {\bf p} \mbox{\boldmath $\sigma$})\phi({\bf r}).
\ee
With this expression, the matrix element takes the form
\be
 M_{fi} = \fr{2m_e}{f_a}(\bar\psi i \gamma_5 \psi)_{fi} = 
\fr{i}{f_a}\langle f | ({\bf p} \mbox{\boldmath $\sigma$})\left(1 - \fr{E_f-U({\bf r})}{2m_e}\right)-
\left(1 - \fr{E_i-U({\bf r})}{2m_e}\right)({\bf p} \mbox{\boldmath $\sigma$})|i\rangle.
\ee
Given the Hamiltonian $H_0 = {\bf p}^2/(2m_e) +U({\bf r})$, with $H_0|i(f)\rangle = E_{i(f)}|i(f)\rangle$,
straightforward quantum mechanical manipulations reduce the matrix element to
\begin{eqnarray}
 M_{fi} &=& -\fr{i}{f_a}\langle f | ({\bf p} \mbox{\boldmath $\sigma$})\fr{E_f-H_0}{2m_e}
- \fr{E_i-H_0}{2m_e}({\bf p} \mbox{\boldmath $\sigma$})|i\rangle \nonumber\\
 &=& -\fr{im_a}{m_e f_a}\langle f | ({\bf p} \mbox{\boldmath $\sigma$})|i\rangle
= \fr{m_a^2}{f_a} \langle f | ({\bf r} \mbox{\boldmath $\sigma$}) |i\rangle,  
\end{eqnarray}
where we have used $E_f-E_i=m_a$.  The result is identical to (\ref{mat_el}) as expected.

\subsection{Vector DM }

{\it Decays}:
The vector model of keV dark matter has an important distinction when compared to 
scalar/pseudoscalar models, as the direct decay to two photons is 
strictly forbidden, regardless of how the new vector particle is introduced in the 
model. 

The decay to three $\gamma$ quanta is allowed at the loop level. In the limit 
$m_V\ll m_e$, the electron coupling will then generate a dimension-eight 
interaction with photons of Euler-Heisenberg form,
\be
 {\cal L}_{\gamma} = \frac{e^3e'}{720\pi^2 m_e^4} \left(14  
F_{\mu \nu}F_{\nu \lambda}F_{\lambda \sigma} V_{\sigma \mu} - 5 ( F_{\mu \nu} F_{\mu \nu})(F_{\lambda \sigma} V_{\lambda \sigma})\right), \label{eh}
\ee
where $F_{\mu \nu}$ is the electromagnetic field strength tensor and $V_{\mu \nu} = \partial_\mu V_\nu - \partial_\nu V_\mu$ is its analog for the field $V_\mu$ of the massive vector. These operators then mediate the dominant 3$\gamma$ decay below the electron threshold, shown in Fig.~1, and we will 
calculate the spectrum and decay rate explicitly. We find that the full Dalitz plot distribution of  photon energies in the rest frame of V is described by
\ba
\frac{{\rm d}\Gamma}{{\rm d} \omega_1 {\rm d} \omega_2} &=& \frac{\alpha^3 \, \alpha' \, m_V^3}{2^5 \, 3^6 \, 5^2 \, \pi^3 \, m_e^8} \, 
\left\{ 287\ (\om_1^4+\om_2^4+\om_3^4) + 538\ ( \om_1^2 \om_2^2+\om_1^2\om_3^2+\om_2^2\om_3^2) \right. \nonumber \\
&& \!\!\!\!\!\!\!\!\!\!\!\!\!\!\!\!\!\!\!\!\!\!\!\!\!\! \left.- 556\, \left [\om_1^3(\om_2+\om_3)+\om_2^3(\om_1+\om_3)+\om_3^3(\om_1+\om_2)-\om_1 \om_2 \om_3 (\om_1+\om_2+\om_3)\right] \right \},
\label{dalitz}
\ea
where $\omega_1$, $\omega_2$ and $\omega_3=m_V-\omega_1-\omega_2$ are the final-state photon energies. Integrating over one of 
the photons yields the inclusive one-photon spectrum in the decay,
\be
\frac{{\rm d} \Gamma}{{\rm d} x} = \frac{\alpha^3 \alpha'}{2^7 3^7 5^3 \pi^3} \, \frac{m_V^9}{m_e^8} \, x^3 \left ( 1715 - 3105 x + \frac{2919}{ 2} x^2 \right ),
\label{one_photon}
\ee
where $x=2 \omega/m_V$, so that the physical region for the dimensionless parameter $x$ ranges from 0 to 1. 
Integrating the one-photon spectrum we finally obtain the total decay rate:
\be
\Gamma = \frac{17 \, \alpha^3 \alpha'}{2^7 3^6 5^3 \pi^3} \, \frac{m_V^9}{m_e^8} \approx \left ( 4.70 \times 10^{-8} \right ) \, \alpha^3 \alpha' \, \frac{m_V^9}{m_e^8}.
\label{v3g_total}
\ee

An immediate application of this result is to provide a constraint on the masses and couplings such that the dark
matter lifetime is at least equal to the age of the universe. We find
\be
\tau_{\rm U} \Ga_{V\rightarrow 3\gamma} \la 1 ~~\Longrightarrow ~~ m_V \, (\alpha')^{1/9} \la 1\, {\rm keV}~.
\ee
Even if the lifetime constraint is satisfied, the 
$\gamma$ background created by $V$ decays can be detected 
as a diffuse cosmological $\gamma$ background, or as an
extra contribution to the $\gamma$ background in our galaxy.

\noindent{\it Emission}: 
In considering emission, we will again focus on Compton-like scattering with electrons, $e+\gamma \rightarrow e + V$, 
which will be relevant in considering
energy loss in stars. In the limit where $m_V$ is small compared to the frequency of the absorbed 
photon, and in the nonrelativistic approximation with respect to the electron, this
process has a standard Thomson-like cross-section:
\be
\label{compton}
 \si_{e\gamma \rightarrow eV} = \frac{8\pi \al \al'}{3m_e^2}. 
\ee 
This results in an energy loss flux (energy/volume/time) from a thermal plasma which we can estimate as
\be
\label{cflux}
 \Ph_{e\gamma\rightarrow eV} = n_\gamma n_e \langle \om \si_{e\gamma \rightarrow eV}\rangle = \left(\frac{2\zeta(3)}{\pi^2}T^3\right) 
 \left(\frac{p_F^3}{3\pi^2}\right)\left(\frac{8\pi^5\al\al'}{90\zeta(3)} \frac{T}{m_e^2}\right),
\ee
to again be compared with various constraints on stellar energetics in the next section. As above, this 
rate has to be modified to account for Boltzmann suppression if $m_V \ga T$. 
However, the formulae (\ref{compton}) and (\ref{cflux}) neglect another possibly 
important effect: the effective suppression 
of $V-\gamma$ mixing due to the dynamical mass of the photon inside plasma. 
This screening means that inside a highly-conducting medium the coupling 
constant $\alpha'$ has to be modified as follows:
\be
\alpha' \to \alpha'_{\rm eff} \simeq \alpha'\times \left(  \fr{m_V^2}{m_V^2
-m_D^2}  \right)^2 ~~\to~~ \alpha'_{\rm eff} \simeq
\fr{\alpha'm_V^4}{m_D^4}  ~~{\rm for}~~ m_V\ll m_D  .
\label{alphaeff}
\ee
For temperatures which are low compared to the electron mass, the effective 
dynamical mass (or plasma frequency) can be taken
as $m_D^2 =4\pi\alpha m_e^{-1}(n_{e}+n_{e^+})  $. In the stellar environments where
$n_{e^+}=0$, and $n_{e}$ is not exceedingly large, this would not imply a
strong suppression of the emission rates 
unless $m_V$ is under a few keV. However, 
for cosmological or SN applications, where $T\ga m_e$, the 
high-temperature effective mass is relevant, $m_D^2 = 4\pi\alpha T^2/3$. The consequent
suppression of all $V$-production rates can then be quite significant. 

To estimate $V$-production during supernovae, we note that it is the coupling of $V$ to 
nucleons, and in particular the coupling to the neutron magnetic moment that leads
to the most important production channel.  This logic is motivated by the large number density of 
nuclear matter in the core of the SN, and by the analogy with the axion case, where coupling to 
nucleons provide more stringent constraints. Furthemore, since the nuclear matter in the core is 
mostly neutrons, it is the neutron electromagnetic formfactors, and magnetic moment in 
particular, that mediate $V$-production. 
Adopting the method of Refs.~\cite{Phillips,OP}, we estimate the 
emission of vectors by factorizing the nucleon elastic scattering cross section $\sigma_{NN}$ 
and the probability of $V$-emission due to the neutron spin flip, 
\be
\label{PhiSN}
 \Ph_{NN\rightarrow NNV} \sim \fr{(\mu_N T)^2}{4\pi^2} 
\sigma_{NN}n_N^2 \left( \fr{T}{m_N}\right)^{1/2} T \alpha'_{\rm eff}(T),
\ee
where $\mu_N = 1.9(4\pi\alpha)^{1/2}/(2m_N) $ is the magnetic moment of the neutron, and 
$n_N$ is the neutron number density. 
Assuming typical temperatures on the order of 10 MeV, 
we notice that for $m_V \la 1$ MeV, the $m_D$ in (\ref{alphaeff}) dominates, and it
is possible to reduce (\ref{PhiSN}) to
\be
\label{phireduced}
\Ph_{NN\rightarrow NNV} \sim 10^{-2} \times \fr{\alpha'}{\alpha}\times \fr{m_V^4 n_N^2\sigma_{NN}}{m_N^{5/2}T^{1/2}},
\ee
which has a strong dependence on $m_V$ but  a rather mild temperature-dependence.  One
observes that the primary advantage of the SN limits, namely an enhancement of the emission rates at large 
temperatures, is completely lost in (\ref{phireduced}) due to the strong suppression 
of $\alpha'_{\rm eff}$. Consequently, we do not find any competitive constraints on the parameter space from
SN physics.\footnote{Additional suppression of (\ref{phireduced}) arises from the large residual chemical potential 
for electrons in SN \cite{kr}.} The inclusion of other $V$-production channels ({\em e.g.} from electrons, positrons,
and other charge carriers) would not change this conclusion.

The cosmological abundance of keV-scale vectors has a thermal and also possibly a nonthermal component. 
Thermal emission of vectors due to the operator  $\kappa F_{\mu\nu}V_{\mu\nu}$ is strongly inhibited 
at high temperatures, and thus the most important threshold for emission is the 
$e^+e^-$ threshold at $T~\sim 0.5$MeV. Production of $V$ occurs via the processes $e+\gamma \rightarrow e + V$
and $e+e^+ \rightarrow V+\gamma$, and even below the 
the electron threshold via $\gamma + \gamma \to \gamma +V$ mediated by a
loop diagram, although the latter process is probably going to be subdominant. 
In this paper we will concentrate on the mass range $m_V \la 100$ keV.
This allows for some additional simplifications, as the production rate is peaked  below $T=m_e$
due to (\ref{alphaeff}). For $m_V=10$ keV, the peak in production occurs at $\sim 150$ keV,
which is predominantly due to Thomson-like scattering off the remaining  
electrons and positrons. This allows for a nonrelativistic treatment of the
electrons, and justifies the use of the Thomson-like formula. 
With these simplifications, we can calculate the freeze-out abundance of $V$-particles per photon
as a function of $m_V$ and $\alpha'$,
\begin{eqnarray}
Y_V(m_V,\alpha') \equiv \fr{n_V}{n_\gamma} \simeq \int_0^{m_e} \fr{(n_{e}+n_{e^+}) 
\langle \sigma_{e\gamma \rightarrow eV} c\rangle dT}{HT}
\simeq \fr{\alpha'}{\alpha}\times \fr{8\pi\alpha^2}{3m_e^2}\int_0^{m_e} \fr{2n_em_V^4dT}{(m_V^2 + m_D^2)^2HT},
\label{YV} 
\end{eqnarray}
where we took into account that the number densities of electrons and positrons are approximately 
equal, $n_{e}\simeq n_{e^+} \simeq 2^{-1/2}(m_eT/\pi)^{3/2}\exp(-m_e/T)$. Once again, we assume that the $V$-sector is not in chemical equilibrium with the SM, 
and the production rate of $V$-particles is slower than the Hubble expansion rate $H$. 
The upper limit of integration is chosen as $\sim m_e$, the scale where the nonrelativistic 
approximation breaks down, but can formally be extended to infinity because the integrand 
has a maximum below $T=m_e$ and falls quickly at high temperatures. 
Using (\ref{YV}), we can immediately calculate the 
corresponding contribution of $V$-particles to the total energy density today:
\be
\Omega_V(m_V,\alpha') = \Omega_{\rm baryon}~ \fr{m_V}{m_N}~\fr{n_\gamma}{n_{\rm baryon}}~Y_V = 
73 ~Y_V\times \fr{m_V}{\rm keV}.
\label{OmegaV}
\ee
Choosing $m_V=10$ keV for example and tuning $\Omega_V$ to the measured value of $\Omega_{DM}$, \newline
i.e. $\Omega_V(10~{\rm keV},\alpha')=0.2$,
we find
\be
\fr{\alpha'}{\alpha} \simeq 10^{-20}.
\ee
This tiny coupling certainly justifies referring to $V$ dark matter as a super-WIMP.

Eqs.~(\ref{YV}) and (\ref{OmegaV}) provide a way of estimating the thermal component of 
$V$ dark matter created at the electron threshold. 
There are, of course, other possiblities for creating additional contributions 
to $\Omega_V$. For example, inflation may end with some inflaton decays to 
particles in the U(1)$'$ sector; higher dimensional operators may provide an efficient 
way of transfering energy from the SM sector into the U(1)$'$ \cite{FP};
the Higgs$'$-strahlung processes, $e^++e^- \rightarrow V^* \rightarrow V+h'$,
might also be important, etc. All of these mechanisms are rather model-dependent, and 
can only {\em add} to $\Omega_V$ on top of the estimates (\ref{YV}) and (\ref{OmegaV}). 

It is worth dwelling on the Higgs$'$-strahlung process for $V$-production, as it  
can be an important mechanism for the following reason. The virtuality of the  
$\gamma-V$ line is $q^2>4 m_e^2$, and thus one does not have any thermal 
 suppression for this process. 
Moreover, the cross section for  Higgs$'$-strahlung remains finite in the $m_V\to 0 $ limit, 
and in our model scales as $\sim \alpha' ({\tilde e})^2/E^2$, where $E$ is the 
center of mass energy, and ${\tilde e}$ is the gauge coupling in the U(1)$'$ sector.
For a fixed value of ${\tilde e}$, the smaller the mass of $m_V$ we consider, the more important 
the Higgs$'$-strahlung production of $V$ will be. In this paper, we choose to ignore it 
noting that for our choice of mass range, $m_V\ga$ keV, the thermal suppression (\ref{alphaeff}) 
at the electron threshold is at most a factor of $O(10)$, and we can 
suppress the Higgs$'$-strahlung relative to the Thomson production of 
$V$ by choosing ${\tilde e}\la e$.\footnote{ We note that in the recent paper \cite{Ringwald},  the cosmological 
abundance of $V$ bosons for $m_V < 1$ eV is computed, but without including the Higgs$'$-strahlung
process which is likely to be very significant for this mass range. 
Indeed, Ref.~\cite{Ringwald} often considers rather large mixing angles $\kappa$,
which would completely populate $V$ and $h'$ right at the electron threshold, 
creating a minimum of four new degrees of freedom,
unless $\kappa {\tilde e}$ is chosen to be less than $\sim 10^{-8}e$.
Having four new degrees of freedom at $T=1$ MeV is now firmly excluded by Big Bang Nucleosynthesis 
constraints. }

\noindent{\it Absorption}:
The absorption of $V$ DM by atoms is very similar to the photoelectric effect 
with photon energy $\omega = m_V$. The only difference is again the factor $\exp(i{\bf kr})$ incorporating the photon spatial momentum. 
As discussed previously in connection with the axioelectric effect this factor can be safely approximated by one. 
Thus, for our estimates it  suffices to take 
\be
\label{sigmaV}
\fr{\sigma_{abs} v}{\sigma_{photo}(\omega=m_V)c} ~ \simeq ~\fr{\alpha'}{\alpha}.
\ee
Converting this cross section into a counting rate 
gives
\be
\label{countingV}
R \simeq   \frac{4 \times 10^{23}}{A} \fr{\alpha'}{\alpha} \left( \fr{\rm keV}{m_V}\right)
\left(\fr{\sigma_{photo}}{\rm bn}\right)\;{\rm kg^{-1}day^{-1}} .
\ee

\section{Direct detection of keV  superWIMPs}

In this section, we use the various interaction rates determined above to
 assess the viability of direct detection of bosonic
keV-scale superWIMP dark matter. We will focus on the photoelectric-type 
ionization cross-section as the primary source of 
a direct detection signal, and also consider how various indirect 
constraints will cut into the available parameter space. In this
regard, we consider energy loss from stars and 
supernovae, and also the $\gamma$-background produced
by decay. We will now consider the pseudoscalar and vector models in turn.

\subsection{Pseudoscalar DM}

As has already been noted above, the monochromatic decay $a\rightarrow \gamma\gamma$ leads to a rather stringent
constraint on $C_\gamma/f_a$ not just through the need to have a sufficiently 
long lifetime, but more significantly through the
induced source of galactic (and also diffuse cosmic) X-rays. 
The constraints on such monochromatic sources in the galaxy are
quite stringent, and for this reason we will adopt the most conservative model, 
Case C, in relating $C_\gamma$ to the scales
$f_a$ and $m_a$. This allows us to restrict our attention to a 2-dimensional 
parameter space, $f_a$ vs $m_a$, which we show in
Fig.~2. The various contours on this plot are described below:

\begin{figure}[t]
\begin{picture}(450,240)
\put(0,0){\centerline{\includegraphics[width=12cm]{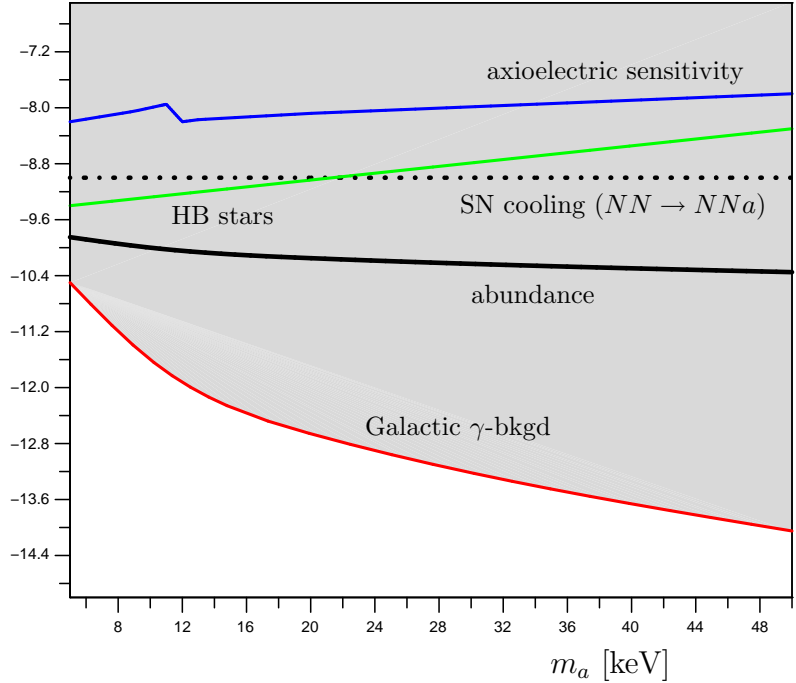}}}
\put(280,0){{$m_a$ [keV]}}
\put(55,270){{$\log({\rm GeV}/f_a)$}}
\put(210,90){\footnotesize {Galactic $\gamma$-bkgd}}
\put(130,170){\footnotesize {~~HB stars}}
\put(256,225){\footnotesize {axioelectric sensitivity}}
\put(246,174){\footnotesize {SN cooling ($NN \rightarrow NNa$)}}
\put(250,140){\footnotesize {abundance}}
\end{picture}
\caption{\footnotesize Fixing $C_\gamma$ in terms of $f_a$ and $m_a$ according to Case $C$, we plot the direct detection sensitivity to pseudoscalar DM
arising from the axioelectric cross-section on Ge, assuming a fiducial sensitivity of the detector equivalent to a 1pb cross-section for a 100 GeV WIMP. We also show the constraints 
arising from the He-burning lifetime in
HB stars, from SN cooling 
via a coupling to the neutron magnetic moment with $f_{aNN}=f_a$, 
and most significantly the monochromatic $\gamma$-background from decays in the Galaxy. The grey
shaded region is excluded by the latter indirect constraints. The thick black line corresponds to the parameters required to reproduce the required dark matter
abundance from thermal production with $f_{abb}=f_a$. }
\end{figure}

\begin{enumerate}
\item The monochromatic $\gamma$-decay leads to a very
strong constraint from observations of the galactic background, and in particular searches for various line sources
in this energy range \cite{gamma}.  For example, the decay width to monochromatic photons in the 10~keV range is constrained to be less than 
approximately $\Ga < 10^{-27}\,s^{-1}$, which we observe is around ten orders of magnitude more stringent than the constraint
on the lifetime. As noted above, the strength of this constraint means that we will only consider Case C in which the photon coupling
is suppressed. The limits obtained in \cite{gamma} can then be used to form an exclusion contour for $f_a$ in terms of $m_a$ as shown in Fig.~2.
The growth of the decay rate with mass is sufficient to render the constraint  more stringent at the upper end of the 
mass range considered. The coupling constants for Cases A and B as well as for
 scalar dark matter are even more strongly constrained. 

\item An important constraint on any new light states that couple to photons or electrons is that the new energy loss mechanisms should
not severely disrupt the life-cycles of stars. Earlier, we used a Compton-like emission process to estimate the energy flux into $a$ particles.
We will consider two constraints, the first of which constrains energy loss from He-burning Horizontal branch (HB) stars in globular clusters. Following Raffelt
\cite{raffelt} (see also Raffelt and Weiss \cite{rw} for constraints from He ignition), this constrains the energy flux to $\Ph < 10^{-42}\,$MeV$^5$ at a 
density of $\rh \sim 10^4$g/cm$^3$ and a temperature of $T\sim 10$~keV. We plot the corresponding exclusion contour in Fig.~2, 
which degrades for $m_a \gg T$ due to the Boltzmann suppression of photons with $\omega > m_a$.
 
 An energy loss constraint which is less sensitive to the mass, at least 
in the keV-range, arises from cooling of the supernova
 core, e.g. in SN1987A. In this case, the constraint on the flux is 
$\Ph < 10^{-14}\,$MeV$^5$ at a density of $\rh \sim 10^{14}$g/cm$^3$ and a 
temperature of $T\sim 30$~MeV \cite{raffelt}. At this core temperature, the $a$ particles are effectively massless and from the electron coupling
we find  a rather weak  horizontal exclusion contour of $f_a > 8 \times 10^7$~GeV. However, allowing for derivative couplings to quarks, in analogy to the 
electron coupling considered above, a stronger constraint on $f_a\sim f_{aNN}$ ensues from the induced coupling to the neutron magnetic moment
in the degenerate SN core leading to the contour shown in Fig.~2.

\item  
We also include a line that corresponds to $\Omega_a = \Omega_{DM}$, using Eq. (\ref{Omegaa}), 
assuming $B_b \sim 0.1$ and choosing $f_{abb}=f_a$. This ``natural abundance" line competes 
with the SN constraints, but is already deeply inside the area excluded by the galactic X-ray 
constraints.

\item The axioelectric sensitivity contour assumes that an experiment like CDMS has a sensitivity 
to the ionization signal from absorption which is equivalent to its sensitivity to the recoil of a 100 GeV
WIMP with a cross-section per nucleus of 1pb. This is simply a benchmark point, and the contour can be
rescaled according to the fact that $\langle \si v\rangle \propto f_a^{-2}$. This sensitivity line is derived 
under the assumption that the $a$-bosons are the dominant component of the dark matter energy density in the 
solar neighborhood, which again can easily be rescaled to a more 
generic case.  Note that we have made use of the well-measured photoelectric cross-section on Ge, which has a 
sharp break at around 11~keV, which is smoothed
out somewhat in the contour.
\end{enumerate}

\subsection{Vector DM}

For vector DM, gauge invariance is rather important in restricting the photon decay rate, and we observe in this model that the indirect constraints
particularly from the $\gamma$-background are considerably weaker. This model has only two parameters, and we plot the 
space $\alpha'/\alpha$ vs $m_V$ in Fig.~3. The various contours are the same as those described for pseudoscalar DM, and use the
same constraints as above. We briefly describe the distinctions below:

\begin{figure}[t]
\begin{picture}(450,240)
\put(0,0){\centerline{\includegraphics[width=12cm]{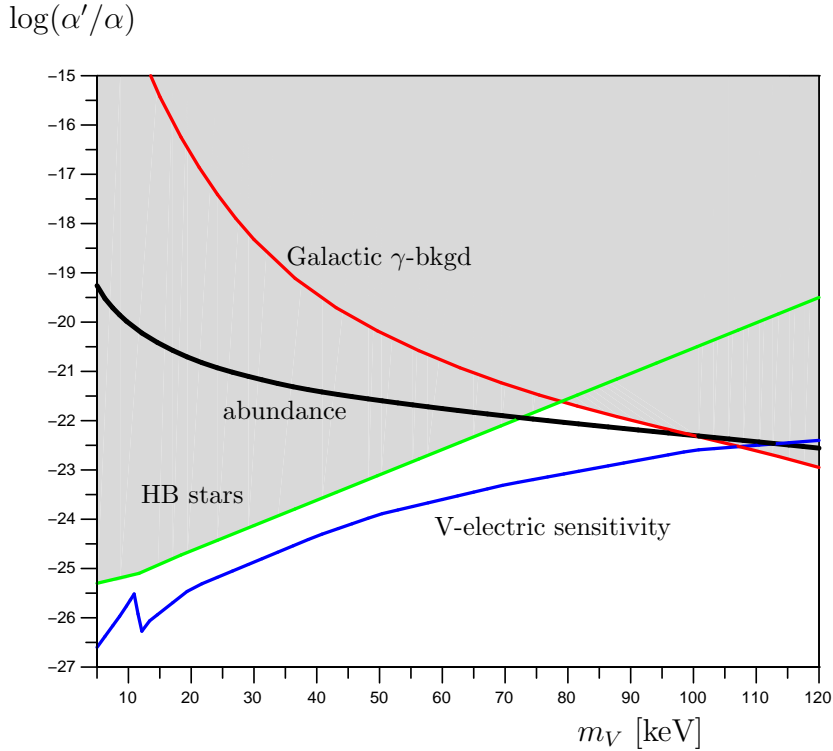}}}
\put(280,0){{$m_V$ [keV]}}
\put(65,270){{$\log(\al'/\al)$}}
\put(170,180){\footnotesize {Galactic $\gamma$-bkgd}}
\put(115,90){\footnotesize {HB stars}}
\put(226,78){\footnotesize {V-electric sensitivity}}
\put(146,122){\footnotesize {abundance}}
\end{picture}
\caption{\footnotesize We plot the direct detection sensitivity to vector 
DM from the V-electric cross-section on Ge, assuming a fiducial sensitivity of the detector equivalent 
to a 1pb cross-section for a 100 GeV WIMP. We also show the constraints 
from the He-burning lifetime in
HB stars, 
and the $\gamma$-background from 3$\gamma$-decays in the Galaxy. The grey shaded region is again excluded by the indirect constraints, while the thick black 
line corresponds to the parameters required to reproduce the required dark matter
abundance from thermal production.}
\end{figure}

\begin{enumerate}
\item The galactic gamma background is less dominant in this case, and the
constraint here is an estimate that comes from relaxing the bound on
monochromatic lines by an order of magnitude to account for the
broader, but still quite peaked, distribution for the 3$\gamma$-decay. The constraint is
again stronger for larger mass.

\item The stellar bounds again arise from energy loss due to the Compton-like processes discussed in the previous
section. We observe that the stellar constraints are particularly strong for $m_V$ of order the core temperature of HB stars\footnote{We thank G. Raffelt for 
emphasizing the importance of constraints from the He-burning lifetime of HB stars in these models.}, but degrades exponentially  for $m_V > 10$ keV
due to Boltzmann suppression,  while in this case SN 
physics does not provide strong constraints anywhere on this plot 
on account of the effective photon mass suppression of  $\alpha'_{\rm eff}$.

\item The V-electric sensitivity was obtained in the same way as for
the pseudoscalar, and we see that in this case the prospects for direct detection look particularly strong, and
indeed this approach may have the best sensitivity by an order of magnitude if the relevant experiments can 
discriminate the ionization signal. As before, this line is calculated assuming that $V$ bosons are the dominant 
component of  galactic dark matter.

\item Remarkably, part of the natural abundance line, calculated using (\ref{YV}) and (\ref{OmegaV}), 
falls within the region allowed by the astrophysics constraints. It can nonetheless be probed rather 
effectively with direct dark matter searches.

\end{enumerate}

\subsection{Discussion of annual modulation}

An annual modulation of the counting rate is a welcome feature in the 
direct detection of WIMPs. In the case of elastic WIMP-nucleus 
scattering, the cross section typically has an $s$-wave component, which
is a constant independent of velocity. The counting rate, however, is 
proportional to the total dark matter flux and thus acquires
a seasonal modulation due to the Earth's motion
at the level of $\Delta v_{DM}/v_{DM} \sim 0.05$. 
The DAMA/NaI and DAMA/Libra collaborations have utilized this idea to 
search for WIMP-nucleus recoils, using the annual modulation as a filter 
to separate signal from background. 

\subsubsection{Suppressed modulation of the super-WIMP counting rate}

In principle, the absorption of bosonic superWIMPs could also generate an
annual modulation of the signal \cite{DAMA1}. This idea, however, immediately
runs aground because inelastic cross sections are typically inversely 
proportional to the velocity of the incoming nonrelativistic particles \cite{LL}. 
In combination with the flux, this renders the counting rate independent of the velocity, 
and therefore un-modulated by the Earth's motion at the experimentally relevant percent level. 
We find that this is indeed the case for both classes of models  considered
in this paper, pseudoscalar and vector superWIMP  dark matter, with rates given in (\ref{countinga}) and 
(\ref{countingV}). We can nonetheless estimate the degree of modulation in 
the absorption, noting that the dark matter velocity
enters  via $\exp(i{\bf kr})$, which in the Born approximation can be 
combined with $\exp(i{\bf pr})$ representing the wavefunction of the outgoing 
photo-electron. Therefore, modulation of the counting rate
arises primarily due to the modulation of $p$. Since $k\ll p$ and the cross section is generally a 
smooth function of energy, we estimate that 
\be
{\rm Modulated~ absorption} ~ \sim \fr{\Delta k}{p} \sim \fr{m_{a(V)}\Delta v_{DM}}{\sqrt{m_e m_{a(V)}}c}
\sim \fr{\Delta v_{DM}}{c} \times  \sqrt{m_{a(V)}/m_e} .
\label{actualmod}
\ee 
For the characteristic energy range of the DAMA signal, $m_{a(V)}\sim 5~{\rm keV}$,
the modulation does not exceed $10^{-5}$. One could potentially worry that $m_{a(V)}$ may turn 
out to be {\em exactly} equal to some ionization threshold where the assumption of a
smooth  cross section as a function of energy breaks down, and the momentum of the 
photoelectron is suppressed relative to its natural value. In this case the 
modulated part of the cross section can be enhanced, but not by the four orders of magnitude 
required to explain the DAMA signal. This is because the detector comprises many-electron atoms and 
unmodulated absorption by other shells would become important. From Eq.~(\ref{actualmod}),
one immediately concludes that the reported DAMA modulation signal cannot be explained
by the absorption of axion-like or vector-like particles.  Specifically, if the couplings are tuned in 
such a way that the modulated part of the axio-electric effect matches the
DAMA signal, the unmodulated part will be in excess of the total number
of events that DAMA observes by four orders of magnitude.\footnote{In the analysis
of the axioelectric process in \cite{DAMA1}, we should note that, besides the
omission of the leading term (\ref{Hint}) in the axion-electron Hamiltonian, the subleading
term $\sim {\bf \sigma k}$ is also incorrectly averaged over the wavefunctions of 
the initial and final electrons. Since the axion  momentum  ${\bf k}$ 
is an external vector, the matrix element of the electron spin operator
between two states with different energies can give a non-zero result only on taking the 
spin-orbit interaction into account, which is not consistent with  \cite{DAMA1}.}

\subsubsection{Solar neutrino backgrounds}

Leaving aside the issue of (un)modulated absorption of superWIMPs, 
we would like to discuss other possible sources of annual modulation in 
ionization. For example, is it possible that emission from the Sun creates 
a modulated ionization signal? Although the photons themselves 
do not reach underground facilities, solar neutrinos can do so and 
 create some amount of ionization. 
The size of the modulation at the level of $\sim 0.03-0.05$ \cite{DAMA2}
is roughly consistent with that of the solar neutrino flux
due to the eccentricity of the Earth's orbit. However, this modulation is expected to be in 
anti-phase with the DAMA result because in the (northern hemisphere) summer the Earth is farther away from the 
Sun than in the winter, and consequently the flux of neutrinos is slightly lower during the 
summer months. Since solar neutrinos have energies much in excess of 
atomic ionization thresholds, one also does not expect a concentration of the 
ionization signal around a few keV, modulo unknown solid state effects.
We can estimate the importance of this (modulated) neutrino background more generally. 
The contribution of neutral currents to elastic scattering on nuclei that mimic recoil events 
has been estimated previously in \cite{nc}. It is only the most energetic fraction of the solar 
neutrino flux that is capable of creating nuclear recoils with more than a 1 keV energy release. 
The rate of such events for DAMA should not exceed $10^{-3}$ kg$^{-1}$day$^{-1}$ \cite{nc}. 
Ionization may also be created by the main part of the $pp$ neutrino flux. 
Our estimates show that the total counting rate due to ionization by $pp$ neutrinos 
is also at the level of $10^{-3}$ kg$^{-1}$day$^{-1}$, providing a modulated 
counting rate that is about two-to-three orders of magnitude smaller than the reported DAMA signal. 
Thus solar neutrinos do not induce a
modulated signal capable of matching the amplitude observed by DAMA.
However, for  future high-sensitivity searches for bosonic superWIMPs, 
of the kind advocated in this paper, 
the ionization created by solar neutrinos may constitute an important source of background. 

To conclude this section, in a more speculative vein one cannot help 
noticing that the energy range for the modulated DAMA signal, $2-6$ keV, claimed in 
\cite{DAMA2} is comparable to the temperature of the solar core, and indeed matches the 
energy range where the emission of exotic massless particles would naturally be peaked \cite{raffelt}. 
This could be ``standard" axions, or massless U(1)$'$ bosons coupled to the SM via marginal 
operators \cite{Bogdan}, or another similar type of exotic. Again, the ionization signal created by these 
exotic particles can have a 3\% annually modulated component, but it will be ``$\pi$-shifted"
relative to the DAMA signal.  Only if the absorption within the Earth is somehow 
an important effect could the integrated day-night effect potentially induce 
modulation with a maximum in June and a minimum in December. We believe that this latter
explanation can be directly checked using the DAMA and DAMA/Libra datasets.

\section{Concluding remarks}

With vast resources now justifiably being devoted toward the direct detection of 
dark matter, it appears all but clear that the scientific scope of these
searches should be diversified. While a characteristic elastic WIMP-nucleus recoil may remain
the main benchmark scenario for these experiments, the detection possibilities for other generic
classes of dark matter  should certainly be exploited. In this paper, we have shown that
bosonic superWIMPs represent a legitimate and feasible target.

The models analyzed in this paper can naturally produce the required relic abundance of dark matter, 
once the couplings $\alpha'$ or $(m_e/f_a)^2$ are in the superweak $\sim O(10^{-20})$ range. 
The feeble nature of this coupling to the SM is partially overcome 
by a factor of $(c/v_{DM}) \times (m_{\rm WIMP}/m_{a(V)})$ which enhances superWIMP absorption 
relative to WIMP scattering and renders direct detection feasible. 
We have analyzed models that do not require any special fine-tunings and enjoy 
protection for the small mass scales due to the symmetries of their interactions:
gauge symmetry for the model of secluded U(1)$'$, and 
shift symmetry for the pseudoscalar model. 

The result of our analysis has revealed that for the pseudoscalar
model, direct detection sensitivity may compete with the red giant, solar and 
SN constraints. However, we found that the model is most strongly 
constrained by limits on monochromatic X-ray lines in our galaxy, that rule out
much of the interesting regions of the parameter space. 
In vector models of dark matter, the production of X-rays is 
strongly inhibited by gauge invariance, and for the model studied here 
direct searches for ionization are apparently capable of probing the 
most interesting range of the coupling--mass parameter space, namely that consistent with 
the observed dark matter energy density. Interestingly, part of this range is not strongly constrained by indirect
astrophysical probes, away from the low mass region where stellar energy loss constraints become
important and the higher mass range where the $\gamma$-background is too large. Finally, contrary to some 
existing claims in the literature,  we have found that the absorption of superWIMPs does not 
lead to an annual modulation of the ionization signal at a level that would
be of experimental interest.

\subsection*{Acknowledgements}

M.P. would like to thank Georg Raffelt for correspondence and for numerous helpful comments following the 
submission of v1 of this paper to the arXiv. We would also like to thank Marieke Postma and Javier Redondo
for pointing out an error in the stellar bound shown on an earlier version of Fig.~3.
The work of A.R. and M.P. is supported in part by NSERC, Canada, and research at the Perimeter Institute
is supported in part by the Government of Canada 
through NSERC and by the Province of Ontario through MEDT. The work of M.V. is supported in part by the DOE grant DE-FG02-94ER40823.


\begin{thebibliography}{99}

\bibitem{review}
{\it see e.g.} G.~Jungman, M.~Kamionkowski and K.~Griest,
  Phys.\ Rept.\  {\bf 267}, 195 (1996)
  [arXiv:hep-ph/9506380];
G.~Bertone, D.~Hooper and J.~Silk,
  Phys.\ Rept.\  {\bf 405}, 279 (2005)
  [arXiv:hep-ph/0404175].
  
  \bi{lw}
  B.~W.~Lee and S.~Weinberg,
  Phys.\ Rev.\ Lett.\  {\bf 39}, 165 (1977);
  M.~I.~Vysotsky, A.~D.~Dolgov and Y.~B.~Zeldovich,
  JETP Lett.\  {\bf 26}, 188 (1977)
  [Pisma Zh.\ Eksp.\ Teor.\ Fiz.\  {\bf 26}, 200 (1977)].
  
      
  \bi{sterile}
  S.~Dodelson and L.~M.~Widrow,
  Phys.\ Rev.\ Lett.\  {\bf 72}, 17 (1994)
  [arXiv:hep-ph/9303287];
  A.~D.~Dolgov and S.~H.~Hansen,
  Astropart.\ Phys.\  {\bf 16}, 339 (2002)
  [arXiv:hep-ph/0009083].
  
  \bi{swimp}
  J.~R.~Ellis, D.~V.~Nanopoulos and S.~Sarkar,
  Nucl.\ Phys.\  B {\bf 259}, 175 (1985);
    J.~R.~Ellis, G.~B.~Gelmini, J.~L.~Lopez, D.~V.~Nanopoulos and S.~Sarkar,
  Nucl.\ Phys.\  B {\bf 373}, 399 (1992);
  J.~L.~Feng, A.~Rajaraman and F.~Takayama,
  Phys.\ Rev.\  D {\bf 68}, 063504 (2003)
  [arXiv:hep-ph/0306024].
  
  \bi{DAMA1}  R.~Bernabei {\it et al.},
  Int.\ J.\ Mod.\ Phys.\  A {\bf 21}, 1445 (2006)
  [arXiv:astro-ph/0511262].

  \bi{cogent} C.~E.~Aalseth {\it et al.},
  arXiv:0807.0879 [astro-ph].
  
    
  \bi{DAMA3}  R.~Bernabei {\it et al.}  [DAMA Collaboration],
  Phys.\ Lett.\  B {\bf 480}, 23 (2000).
  
  
  \bi{DAMA2} R.~Bernabei {\it et al.}  [DAMA Collaboration],
  arXiv:0804.2741 [astro-ph].


\bi{dama_wimp}
J.~L.~Feng, J.~Kumar and L.~E.~Strigari,
  arXiv:0806.3746 [hep-ph];
F.~Petriello and K.~M.~Zurek,
  arXiv:0806.3989 [hep-ph];
S.~Chang, G.~D.~Kribs, D.~Tucker-Smith and N.~Weiner,
  arXiv:0807.2250 [hep-ph].

  
  \bi{PRV}  M.~Pospelov, A.~Ritz and M.~B.~Voloshin,
  Phys.\ Lett.\  B {\bf 662}, 53 (2008)
  [arXiv:0711.4866 [hep-ph]].
  

 \bi{cdms}
D.~S.~Akerib {\it et al.}  [CDMS Collaboration],
  Phys.\ Rev.\ Lett.\  {\bf 96}, 011302 (2006)
  [arXiv:astro-ph/0509259];
  Z.~Ahmed {\it et al.}  [CDMS Collaboration],
  arXiv:0802.3530 [astro-ph].
    
  \bi{xe}
  J.~Angle {\it et al.}  [XENON Collaboration],
  arXiv:0706.0039 [astro-ph].
  

\bi{Uprime}  W.~F.~Chang, J.~N.~Ng and J.~M.~S.~Wu,
  Phys.\ Rev.\  D {\bf 74}, 095005 (2006)
  [arXiv:hep-ph/0608068]; W.~F.~Chang, J.~N.~Ng and J.~M.~S.~Wu,
  Phys.\ Rev.\  D {\bf 75}, 115016 (2007)
  [arXiv:hep-ph/0701254].


\bi{Ringwald} J.~Jaeckel, J.~Redondo and A.~Ringwald,
  arXiv:0804.4157 [astro-ph].

  
  \bibitem{Fayet}
  P.~Fayet,
  Nucl.\ Phys.\  B {\bf 347}, 743 (1990).

\bi{Langacker}  P.~Langacker,
  arXiv:0801.1345 [hep-ph].
  
  \bi{valle}
  F.~Bazzocchi, M.~Lattanzi, S.~Riemer-Sorensen and J.~W.~F.~Valle,
  JCAP 08, 013 (2008)
  [arXiv:0805.2372 [astro-ph]].
  
 \bi{rw}
  G.~Raffelt and A.~Weiss,
  Phys.\ Rev.\  D {\bf 51}, 1495 (1995)
  [arXiv:hep-ph/9410205].
  
  
\bi{raffelt}
G.~G.~Raffelt,
  Ann.\ Rev.\ Nucl.\ Part.\ Sci.\  {\bf 49}, 163 (1999)
  [arXiv:hep-ph/9903472].
  
  
  \bi{Dimopoulos} F.~T.~Avignone {\it et al.},
  Phys.\ Rev.\  D {\bf 35}, 2752 (1987); S.~Dimopoulos, G.~D.~Starkman and B.~W.~Lynn,
  Phys.\ Lett.\  B {\bf 168}, 145 (1986).
  
  
  
  \bi{Phillips} C.~Hanhart, J.~A.~Pons, D.~R.~Phillips and S.~Reddy,
  Phys.\ Lett.\  B {\bf 509}, 1 (2001)
  [arXiv:astro-ph/0102063].
  
  \bibitem{OP}
  K.~A.~Olive and M.~Pospelov,
  Phys.\ Rev.\  D {\bf 77}, 043524 (2008)
  [arXiv:0709.3825 [hep-ph]].
  
  \bi{kr}A.~Kopf and G.~Raffelt,
  Phys.\ Rev.\  D {\bf 57}, 3235 (1998)
  [arXiv:astro-ph/9711196].

  
  
  \bi{FP} A.~E.~Faraggi and M.~Pospelov,
  Astropart.\ Phys.\  {\bf 16}, 451 (2002)
  [arXiv:hep-ph/0008223].


\bi{gamma}
 O.~Ruchayskiy,
  arXiv:0704.3215 [astro-ph];
  H.~Yuksel, J.~F.~Beacom and C.~R.~Watson,
  arXiv:0706.4084 [astro-ph];
   A.~Boyarsky, D.~Iakubovskyi, O.~Ruchayskiy and V.~Savchenko,
  arXiv:0709.2301 [astro-ph];
A.~Boyarsky, D.~Malyshev, A.~Neronov and O.~Ruchayskiy,
  arXiv:0710.4922 [astro-ph].

  

  
   \bi{LL} L.~D.~Landau and E.~M.~Lifshits, {\it Quantum Mechanics (Non-relativistic
Theory)}, Third Edition, Pergamon, Oxford, 1977.

  
  \bi{nc} J.~Monroe and P.~Fisher,
  Phys.\ Rev.\  D {\bf 76}, 033007 (2007)
  [arXiv:0706.3019 [astro-ph]].
  
  
  \bi{Bogdan}  B.~A.~Dobrescu,
  Phys.\ Rev.\ Lett.\  {\bf 94}, 151802 (2005)
  [arXiv:hep-ph/0411004].


  \end{thebibliography}
\end{document}